\documentclass[12pt]{article}

\usepackage{amssymb}
\usepackage{amsmath}
\usepackage{epsfig}

\begin{document}

\title{\bf Kinematic Description of Ricci Solitons in Fluid Spacetimes}

\author{Umber Sheikh\thanks{umbersheikh@gmail.com}
\\Department of Applied Sciences, National Textile University,
\\Faisalabad-37610, Pakistan.
}

\date{}
\maketitle

\begin{abstract}
We consider the kinematics of specific fluid spacetimes
admitting timelike congruences of Ricci Solitons. These fluids
includes string cloud, string fluid, perfect fluid,
radially symmetric fluid, anisotropic fluid and
relativistic magneto-fluid. Results are obtained and
important physical aspects are discussed.
\end{abstract}


\section{Introduction}

The concepts of symmetry and beauty are interrelated.
In-fact, invariants are the deep truths, i.e., things that
do not change. Invariants are defined by symmetries which
in turn define properties of nature are conserved. Selfsame
symmetries like conformal collineations, Ricci and matter
inheritance collineations etc., are those which appeal to
the senses in art and natural forms like snowflakes and
galaxies. Thus, the fundamental truths are raised on basis
of symmetries which contains deep kind of beauty in them.

The symmetries are a part of geometry and thus reveals the physics.
According to Wheeler, \textbf{Spacetime tells
matter how to move; matter tells spacetime how to curve}. There are many
symmetries regarding the spacetime geometry and matter. The metric
symmetries are important as they simplify solutions to many problems.
Their main application in general relativity is that they classify
solutions of Einstein Field Equations. One of these symmetries is
Ricci Solitons associated with the Ricci flow of spacetime geometry.

Ricci flow is a process of deformation of metric in directions of its
Ricci curvature. The Ricci flow can be defined by
following geometric evolution equation \cite{1}
\begin{equation}{\setcounter{equation}{1}}
\partial_t g_{ij}=-2R_{ij}\label{1}
\end{equation}
The stretch or contraction of metric depends on positive or negative
Ricci curvature. The faster deformation of the metric depends upon
the strength of curvature.

The most interesting problem in Ricci flow is to determine its
self similar solutions of Ricci flow namely \textbf{Ricci Solitons}.
These are natural generalization of Einstein metrics \cite{Cao}.
Ricci solitons are fixed points of Ricci flow in a dynamical system.
These are the self similar solutions to the Ricci flow. Ricci solitons
model the formulation of singularities in the Ricci flow.
It is worth mentioning here that the mathematical notion of Ricci
soliton should not be confused with the notion of soliton solutions,
which arise in several areas of mathematical and theoretical physics
and its applications.

Ricci flow is important because it can help in understanding the concepts
of energy and entropy in general relativity. The most interesting property
of the Ricci flow is its tendency to lose memory of initial conditions. It
leads to the property that the irregularities of metric can be evolved
into some regular form. This property is the same as that
of heat equation due to which an isolated system loses the heat for a
thermal equilibrium. Ricci solitons are the points at which the curvature
obeys a self similarity.

Most of recent literature is found regarding the geometry of Ricci
flow and Ricci Solitons. Hamilton \cite{H} was the frontier to introduce
Ricci flow. Ivey \cite{Ivey} discussed the geometry of Ricci Solitons
on compact manifolds. Catino et al. \cite{Catino2} found that complete
gradient expanding Ricci soliton with nonnegative Ricci curvature is
isometric to a quotient of the three dimensional Gaussian soliton.
Munteanu \cite{Mun} discussed the curvature behavior of four
dimensional shrinking gradient Ricci solitons. Kroencke \cite{Kr}
showed that though complex projective spaces of even complex dimension
have infinitesimal solitonic deformations, they are rigid as Ricci
solitons. Catino et al. \cite{Catino3} provided some necessary
integrability conditions for the existence of gradient Ricci solitons in
conformal Einstein manifolds. Catino et al. \cite{Catino4}
proved some of the classification results for generic shrinking
Ricci solitons. Kroencke \cite{Kr2} discussed the stability
of Ricci solitons with respect to Perelman’s shrinker entropy.
Jablonski \cite{J} proved that homogeneous Ricci soliton spaces
are algebraic spaces. Wei and Wu \cite{Wei} discussed the conditions
for Euclidean volume growth of Ricci solitons. Woolgar \cite{Woo}
considered some applications of Ricci flows in static metrics of
general relativity. He also take Ricci solitons in context. Akbar
and Wooglar \cite{Ak} showed an explicit family of complete expanding
solitons as a Ricci flow for a complete Lorentzian metric.

This article is devoted to find the kinematics of spacetimes
admitting Ricci solitons. Section 2 comprises of matter tensors
of different fluid spacetimes and respective Ricci tensors. Section
3 consists of basic kinematics regarding to the congruences in fluids.
The same section contains basic information of Ricci solitons as
spacetime congruences. Section 4 contains the mathematical expressions
modeling the kinematical properties of specific fluid spacetimes
admitting Ricci solitons. Section 5 contains the summary of results.

\section{Matter Tensors of Different Fluid Spacetimes}

The famous Einstein field equations relate geometry of spacetimes
with their physics via following expression:
\begin{equation}{\setcounter{equation}{1}}
G_{ab}\equiv R_{ab}-\frac{1}{2}Rg_{ab}=\kappa T_{ab}
\end{equation}
where $G_{ab}$ is the Einstein tensor.
Substituting $\kappa=1$
\begin{equation}\label{EFE}
R_{ab}-\frac{1}{2}Rg_{ab}=T_{ab}
\end{equation}
The Einstein tensor is entirely made up of Ricci tensor components
or containing the geometry of spacetime. The matter tensor describes
the physics of spacetime.

Consider a fluid spacetime admitting the unit spacelike vector $x_a$
perpendicular to unit timelike vector $u^a.$ Then, $u_au^a=-1,~x_ax^a=1,~ u^ax_a=0.$
The signature of the spacetime be (+,-,-,-). Let $T_{ab}$ be the matter
tensor of this fluid spacetime. Using Eq.(\ref{EFE}) we can evaluate
the Ricci tensor as
\begin{equation}
R_{ab}=T_{ab}-\frac{1}{2}Tg_{ab} \label{RT}
\end{equation}
where $T=T^a_a$ is the trace of matter tensor.

The matter tensor for the string cloud can be written as
\begin{equation}\label{SCM}
T_{ab}=\rho u_au_b-\lambda x_ax_b
\end{equation}
where $\rho$ is the density of the fluid and $\lambda$ is string
tension. Eq.(\ref{RT}) provides us
\begin{equation}\label{SCR}
R_{ab}=\rho u_au_b-\lambda x_ax_b+\frac{g_{ab}}{2}(\rho+\lambda).
\end{equation}

The matter tensor for the string fluid can be expressed as
\begin{equation}
T_{ab}=(\rho+q)(u_au_b-x_ax_b)+qg_{ab}
\end{equation}
where $\rho$ is density of the fluid and $q$ is the pressure on the strings.
Eq.(\ref{RT}) gives
\begin{equation}
R_{ab}=(\rho+q)(u_au_b-x_ax_b)+\rho g_{ab}.
\end{equation}

The matter tensor for perfect fluid can be written as
\begin{equation}
T_{ab}=(\rho+p)u_au_b+pg_{ab}
\end{equation}
where $\rho$ is density of the fluid and $p$ is the pressure.
The use of Eq.(\ref{RT}) leads to
\begin{equation}
R_{ab}=(\rho+p)(u_au_b-x_ax_b)+\frac{1}{2}(\rho-p)g_{ab}.
\end{equation}

The matter tensor for anisotropic fluid can be expressed as
\begin{equation}
T_{ab}=\rho u_au_b+px_ax_b
\end{equation}
where $\rho$ is density of the fluid, $p$ is principal pressure
and transverse pressures measured in orthogonal direction to $x^a$
are all vanishing. Its Ricci tensor can be written as
\begin{equation}
R_{ab}=\rho u_au_b+px_ax_b+\frac{1}{2}(\rho-p)g_{ab}.
\end{equation}

The matter tensor for imperfect fluid is
\begin{equation}
T_{ab}=(\rho+p)u_au_b+pg_{ab}+qx_ax_b
\end{equation}
where $\rho$ and $p$ are respectively density and pressure of the fluid,
and $q$ is the shear of the fluid in $x$-direction. The Ricci tensor for
this fluid is
\begin{equation}
R_{ab}=\rho u_au_b+qx_ax_b+\frac{1}{2}(\rho-p-q)g_{ab}.
\end{equation}

The matter tensor for relativistic magneto-fluid can be written as
\begin{equation}
T_{ab}=(\rho+p)u_au_b+pg_{ab}+\mu\left\{|b|^2\left(u_au_b+\frac{1}{2}g_{ab}\right)-b_ab_b\right\}
\end{equation}
where $\rho$ is the density of fluid, $p$ is the pressure, $\mu$ is
the magnetic permeability and $b_a$ is magnetic flux vector such that
$u^ab_a=0.$ Since, the magnetic field is spacelike, therefore
$b_a=|b|x_a.$ The strength of magnetic field can be determined by
$b^ab_a=|b|^2=B.$ Eq.(\ref{RT}) implies the following Ricci tensor
\begin{equation}
R_{ab}=(\mu B+\rho+p)u_au_b+qx_ax_b+\frac{1}{2}(\mu B+\rho-p)g_{ab}-\mu b_ab_b.
\end{equation}

\section{Basic Kinematics of Spacetime Congruences}

Consider the decomposition of a timelike tidal tensor
\begin{equation}
u_{a;b}=\frac{\theta}{3}h_{ab}+\sigma_{ab}+\omega_{ab}-\dot{u}_au_b.
\end{equation}
Here $\dot{u}_a$ is the acceleration vector of the flow. The projective
tensor $h_{ab}=g_{ab}+u_au_b$ projects a tangent vector perpendicular to
$u^a.$ The rate of separation of flowlines, from a timelike curve tangent
to $u_{a}$ can be expressed by expansion tensor $\theta_{ab}=h^c_ah^d_bu_{(c;d)}.$
The volume expansion of the flow lines is $\theta=h^{ab}\theta_{ab}.$ The
shear between the flow lines can be measured by
\begin{eqnarray}
\sigma_{ab}=\theta_{ab}-\frac{\theta}{3}h_{ab}\nonumber\\\label{Sig}
\Rightarrow \theta_{ab}=\sigma_{ab}+\frac{\theta}{3}h_{ab}.
\end{eqnarray}
The vorticity tensor $\omega_{ab}$ and expansion
tensor $\theta_{ab}$ added to result in
$$u_{a;b}+\dot{u}_au_b=\theta_{ab}+\omega_{ab}$$
where $\omega_{ab}=h^c_ah^d_bu_{[c;d]}.$

For any Riemann manifold, the Ricci Soliton Eq.(\ref{1}) can be written as
\begin{equation}{\setcounter{equation}{1}}
R_{ab}-\frac{1}{2}\pounds _{\xi}g_{ab}=kg_{ab}
\end{equation}
It can also be written as
\begin{equation}
L_{\xi}g_{ab}=2(R_{ab}-kg_{ab}).
\end{equation}
The solitons may be
\begin{itemize}
  \item shrinking for $k>0,$
  \item steady for $k=0,$
  \item expanding for $k>0.$
\end{itemize}

The component form of the equation becomes
\begin{equation}\label{RC}
\xi_{a;b}+\xi_{b;a}=2(R_{ab}-kg_{ab}).
\end{equation}
If $\xi$ is taken to be a timelike vector, then $\xi^a=\xi u^a$ where
$u^a$ is a timelike unit vector. Thus, the above equation becomes
\begin{equation}\label{2}
u_{a;b}+u_{b;a}+(\ln\xi)_{;a}u_b+(\ln\xi)_{;b}u_a=2\xi^{-1}(R_{ab}-kg_{ab}).
\end{equation}

\section{Properties of Fluid Spacetimes Admitting Ricci Solitons}

The aim of this research is to identify the kinematical properties
of different spacetime fluids containing Ricci Solitons. In this respect,
we have calculated some restrictions and specifications on such fluid
spacetimes for the existence of Ricci solitons.
\\
\\
\textbf{Theorem 1:}\\
\\
A $4$-dimensional string cloud spacetime possesses a timelike
Ricci soliton $\xi,$ parallel to fluid flow velocity vector field
$u,$ i.e.,
$\xi_a=\xi u_a,\xi>0$  iff
\begin{description}
  \item[C1] \begin{equation}{\setcounter{equation}{1}}\label{RS11}(\ln\xi\dot{)}=\xi^{-1}\left(\frac{\lambda-\rho}{2}-k\right)\end{equation}
  \item[C2] \begin{equation}\label{RS12}\dot{u}_a=(\ln\xi)_{;a}+(\ln\xi\dot{)}u_a=(\ln\xi)_{;a}+\xi^{-1}\left(\frac{\lambda-\rho}{2}-k\right)\end{equation}
  \item[C3] \begin{equation}\label{RS13}\theta=\frac{3\xi^{-1}}{2}\left(\rho+\frac{\lambda}{3}-2k\right)\end{equation}
  \item[C4] \begin{equation}\label{RS14}\sigma_{ab}=\xi^{-1}\lambda\left(\frac{1}{3}h_{ab}-x_ax_b\right)\end{equation}
  \item[C5] \begin{equation}\label{RS15}\theta_{ab}=\xi^{-1}\left\{\left(\frac{\lambda}{6}-\frac{\rho}{2}+k\right)h_{ab}-\lambda x_ax_b\right\}\end{equation}
  \item[C6] \begin{equation}\label{RS16}\omega_{ab}=\xi^{-1}\left\{\left(\rho+\frac{\lambda}{3}\right)h_{ab}-2kg_{ab}\right\}\end{equation}
\end{description}
\textbf{Proof:}\\
\\
The Ricci soliton equation Eq.(\ref{2}) for a string cloud becomes
\begin{equation}\label{RS1}
u_{a;b}+u_{b;a}+(\ln\xi)_{;a}u_b+(\ln\xi)_{;b}u_a=2\xi^{-1}\left\{\rho u_au_b-\lambda x_ax_b+g_{ab}\left(\frac{\rho+\lambda}{2}-k\right)\right\}.
\end{equation}
\begin{description}
  \item[C1] Contracting Eq.(\ref{RS1})with $u^au^b$ gives Eq.(\ref{RS11}).
  \item[C2] Contracting Eq.(\ref{RS1})with $u^ah^b_c$ we have
  \begin{eqnarray}\label{3}
  \dot{u}_c=(\ln\xi)_{;c}+(\ln\xi\dot{)}u_c.
  \end{eqnarray}
  Substituting the value of $(\ln\xi\dot{)}$ from Eq(\ref{RS11}) gives Eq.(\ref{RS12}).
  \item[C3] Contracting Eq.(\ref{RS1}) with $h^{ab}$ gives Eq.(\ref{RS13}).
  \item[C4] Contracting Eq.(\ref{RS1}) with $h^a_ch^b_d-\frac{1}{3}h^{ab}h_{cd}$ gives Eq.(\ref{RS14}).
  \item[C5] Substituting values from Eq.(\ref{RS13}) and Eq.(\ref{RS14}) in Eq.(\ref{Sig}) gives Eq.(\ref{RS15}).
  \item[C6] The left hand side of Eq.(\ref{RC}) can be expressed as
\begin{equation}
\xi_{a;b}+\xi_{b;a}=\xi(u_{a;b}+u_{b;a}+(\ln\xi)_{;a}u_b+(\ln\xi)_{;b}u_a)\nonumber\\
\end{equation}
Putting value of $\dot{u}_c$ from Eq.(\ref{3}) gives
\begin{eqnarray}
\xi_{a;b}+\xi_{b;a}&=&\xi(u_{a;b}+u_{b;a}+\dot{u}_{a}u_{b}+\dot{u}_{b}u_{a}-2(\ln\xi\dot{)}u_au_b\nonumber\\
&=&2\xi(\theta_{ab}+\omega_{ab}-(\ln\xi\dot{)}u_au_b)\nonumber\\
\Rightarrow 2(\theta_{ab}+\omega_{ab})&=&\xi^{-1}(\xi_{a;b}+\xi_{b;a})+2(\ln\xi\dot{)}u_au_b\nonumber\\
\Rightarrow 2\omega_{ab}&=&\xi^{-1}(\xi_{a;b}+\xi_{b;a})+2(\ln\xi\dot{)}u_au_b-2\theta_{ab}\nonumber\\\label{Vor}
\Rightarrow \omega_{ab}&=&\xi^{-1}(R_{ab}-kg_{ab})+(\ln\xi\dot{)}u_au_b-\theta_{ab}
\end{eqnarray}
Substituting values from Eqs.(\ref{SCR}),(\ref{RS11}) and (\ref{RS15}),
the vorticity tensor for string cloud becomes
$$\omega_{ab}=\xi^{-1}\left[\left(\rho+\frac{\lambda}{3}\right)h_{ab}-2kg_{ab}\right]$$
\end{description}

Substituting values from Eqs.(4.1)-(\ref{RS16}) satisfies Eq.(\ref{RS1}).

In the similar manner, the kinematic quantities related to string fluid,
perfect fluid, anisotropic fluid, imperfect fluid and relativistic
magnetofluid can be obtained.
\\
\\
\textbf{Theorem 2:}\\
\\
A $4$-dimensional string fluid spacetime possesses a timelike Ricci
soliton $\xi,$ parallel to fluid flow velocity vector field $u,$ i.e.,
$\xi_a=\xi u_a,\xi>0$  iff
\begin{description}
  \item[C1] \begin{equation}\label{RS21}(\ln\xi\dot{)}=-\xi^{-1}(k+q)\end{equation}
  \item[C2] \begin{equation}\label{RS22}\dot{u}_a=(\ln\xi)_{;a}+(\ln\xi\dot{)}u_a=(\ln\xi)_{;a}-\xi^{-1}(k+q)u_a\end{equation}
  \item[C3] \begin{equation}\label{RS23}\theta=\xi^{-1}(2\rho-q-3k)\end{equation}
  \item[C4] \begin{equation}\label{RS24}\sigma_{ab}=\xi^{-1}(\rho+q)\left(\frac{1}{3}h_{ab}-x_ax_b\right)\end{equation}
  \item[C5] \begin{equation}\label{RS25}\theta_{ab}=\xi^{-1}\{(\rho-k)h_{ab}-(\rho+q)x_ax_b\}\end{equation}
  \item[C6] \begin{equation}\label{RS26}\omega_{ab}=0\end{equation}
\end{description}
\textbf{Theorem 3:}\\
\\
A $4$-dimensional perfect fluid spacetime possesses a timelike Ricci soliton $\xi,$ parallel to the fluid velocity flow vector field $u,$ i.e.,
$\xi_a=\xi u_a,\xi>0$ iff
\begin{description}
  \item[C1] \begin{equation}\label{RS31}(\ln\xi\dot{)}=-\xi^{-1}\left(\frac{\rho+3p}{2}+k\right),\end{equation}
  \item[C2] \begin{equation}\label{RS32}\dot{u}_a=(\ln\xi)_{;a}-\xi^{-1}\left(\frac{\rho+3p}{2}+k\right)u_a,\end{equation}
  \item[C3] \begin{equation}\label{RS33}\theta=3\xi^{-1}\left(\frac{\rho-p}{2}-k\right),\end{equation}
  \item[C4] \begin{equation}\label{RS34}\sigma_{ab}=0,\end{equation}
  \item[C5] \begin{equation}\label{RS35}\theta_{ab}=\xi^{-1}\left(\frac{\rho-p}{2}-k\right)h_{ab},\end{equation}
  \item[C6] \begin{equation}\label{RS36}\omega_{ab}=0.\end{equation}
\end{description}
\textbf{Theorem 4:}\\
\\
A $4$-dimensional anisotropic fluid spacetime possesses a timelike Ricci soliton $\xi,$ parallel to fluid flow velocity vector field $u,$ i.e.,
$\xi_a=\xi u_a,\xi>0$ iff
\begin{description}
  \item[C1] \begin{equation}\label{RS41}(\ln\xi\dot{)}=-\xi^{-1}\left(k+\frac{\rho+p}{2}\right),\end{equation}
  \item[C2] \begin{equation}\label{RS42}\dot{u}_a=(\ln\xi)_{;a}-\xi^{-1}\left(k+\frac{\rho+p}{2}\right)u_a,\end{equation}
  \item[C3] \begin{equation}\label{RS43}\theta=\xi^{-1}\left(\frac{3\rho}{2}-\frac{p}{2}-3k\right),\end{equation}
  \item[C4] \begin{equation}\label{RS44}\sigma_{ab}=\xi^{-1}p\left(x_ax_b-\frac{1}{3}h_{ab}\right),\end{equation}
  \item[C5] \begin{equation}\label{RS45}\theta_{ab}=\xi^{-1}\left\{\left(\frac{\rho-p}{2}-k\right)h_{ab}+px_ax_b\right\},\end{equation}
  \item[C6] \begin{equation}\label{RS46}\omega_{ab}=0.\end{equation}
\end{description}
\textbf{Theorem 5:}\\
\\
A $4$-dimensional imperfect fluid spacetime possesses a timelike Ricci soliton $\xi,$ parallel to fluid flow velocity vector field $u,$ i.e.,
$\xi_a=\xi u_a,\xi>0$  iff
\begin{description}
  \item[C1] \begin{equation}\label{RS51}(\ln\xi\dot{)}=-\xi^{-1}\left(\frac{\rho+p+q}{2}+k\right),\end{equation}
  \item[C2] \begin{equation}\label{RS52}\dot{u}_a=(\ln\xi)_{;a}-\xi^{-1}\left(\frac{\rho+p+q}{2}+k\right)u_a,\end{equation}
  \item[C3] \begin{equation}\label{RS53}\theta=\xi^{-1}\left(\frac{3\rho-3p-q}{2}+2k\right),\end{equation}
  \item[C4] \begin{equation}\label{RS54}\sigma_{ab}=-\xi^{-1}q\left(\frac{1}{3}h_{ab}-x_ax_b\right),\end{equation}
  \item[C5] \begin{equation}\label{RS55}\theta_{ab}=\xi^{-1}\left\{\left(\frac{\rho-q-p}{2}+\frac{2k}{3}\right)h_{ab}+qx_ax_b\right\},\end{equation}
  \item[C6] \begin{equation}\label{RS56}\omega_{ab}=\frac{-5}{3}k\xi^{-1}h_{ab}.\end{equation}
\end{description}
\textbf{Theorem 6:}\\
\\
A $4$-dimensional relativistic magneto-fluid spacetime possesses a timelike Ricci soliton $\xi,$ parallel to fluid flow velocity vector field $u,$ i.e.,
$\xi_a=\xi u_a,\xi>0$  iff
\begin{description}
  \item[C1] \begin{equation}\label{RS61}(\ln\xi\dot{)}=-\xi^{-1}\left(\frac{\mu B+\rho+3p}{2}+k\right),\end{equation}
  \item[C2] \begin{equation}\label{RS62}\dot{u}_a=(\ln\xi)_{;a}-\xi^{-1}\left(\frac{\mu B+\rho+3p}{2}+k\right)u_a,\end{equation}
  \item[C3] \begin{equation}\label{RS63}\theta=\xi^{-1}\left(\frac{3\rho-3p+\mu B}{2}-3k\right),\end{equation}
  \item[C4] \begin{equation}\label{RS64}\sigma_{ab}=\xi^{-1}\mu B\left(\frac{1}{3}h_{ab}-x_ax_b\right)=\xi^{-1}\mu \left(B\frac{1}{3}h_{ab}-b_ab_b\right),\end{equation}
  \item[C5] \begin{equation}\label{RS65}\theta_{ab}=\xi^{-1}\left\{\left(\frac{\mu B+\rho-p}{2}-k\right)h_{ab}-\mu b_ab_b\right\},\end{equation}
  \item[C6] \begin{equation}\label{RS66}\omega_{ab}=0.\end{equation}
\end{description}

\section{Summary}

A fluid spacetime admitting a timelike Ricci soliton $\xi^a=\xi u^a$ can posses some kinematic properties. The specifications obtained may be summerized as under:
\begin{itemize}
  \item In all the mentioned fluid spacetimes, the timelike congruences have an acceleration vector dependent on time derivative of natural logarithm of $\xi$ as $$\dot{u}_a=(\ln\xi)_{;a}+(\ln\xi\dot{)}u_a.$$ Thus, the fluid exert a net force proportional to the acceleration.
  \item In string cloud spacetine, the shear tensor of Ricci soliton congruences depend on the tension of strings.
  \item The string fluid admits a Ricci soliton congruence with a shear tensor dependent on density of the strings as well as the pressure between the strings.
  \item The Ricci soliton flow lines in a perfect fluid do not possess any shear.
  \item For anisotropic fluid, the shear tensor components of Ricci soliton congruences are parallel pressure dependent whereas in the case of imperfect fluid, these are viscosity dependent.
  \item In relativistic magnetofluid, the shear of Ricci soliton congruences is dependent on magnetic field strength and magnetic permeability.
  \item A perfect fluid admitting a Ricci soliton congruence has volume expansion proportional to its expansion tensor.
  \item In cases of string fluid, perfect fluid and relativistic magnetofluid, the congruences of Ricci solitons have no vorticity, i.e, the congruences have no common point of rotation.
  \item The vorticity tensor of an imperfect fluid is totally dependent on $k,$ i.e, in the case of shrinking solitons the vorticity is negative, for steady solitons there is no vorticity in the congruences whereas expanding solitons admit positive vorticity.
  \end{itemize}

\end{document}